\documentclass{PoS}
\usepackage{amssymb}
\title{Unveiling The Physics of Star Formation and Feedback in Galaxies}

\ShortTitle{Star Formation and Feedback}

\author{\speaker{Fatemeh S. Tabatabaei}$^{1,2}$, Jonathan Braine$^{3}$, Carsten Kramer$^{4}$, Eva Schinnerer$^{5}$, John Beckman$^{1,2,6}$, and Johan Knapen$^{1,2}$ \\
          $^1$Instituto de Astrof\'isica de Canarias,  V\'ia L\'actea S/N, E-38205 La Laguna, Spain \\
          $^2$Departamento de Astrof\'isica, Universidad de La Laguna, E-38206 La Laguna, Spain\\
          $^3$Laboratoire d\'Astrophysique de Bordeaux, 33271 Floirac, France\\
          $^4$Instituto Radioastronomia Milimetrica, 18012 Granada, Spain\\
          $^5$Max-Planck-Institut f\"ur Astronomie, K\"onigstuhl 17, 69117 Heidelberg, Germany\\
          $^6$Consejo Superior de Investigaciones Cient\'ificas\\
          
          E-mail: \email{ftaba@iac.es}}

\abstract{Recent studies show the importance of feedback in the 
evolution of the star formation rate in the Universe. However, 
the nature and physics of the feedback are still pressing questions. 
Radio continuum observations can provide unique dust-unbiased tracers 
of massive star formation and of the interstellar medium (ISM) and 
hence are ideal to address  the regulation of star formation in galaxies. 
Our multi-frequency and multi-resolution radio surveys in nearby galaxies 
enable us to trace various phases of star formation and dissect the thermal and nonthermal ISM in galaxies. 
Mapping the cosmic ray electron energy index and magnetic field strength, we have found observational evidence 
that massive star formation significantly affects the energy balance in the ISM through the injection 
and acceleration of cosmic rays and the amplification of magnetic fields. How the next generation of stars could form in such a magnetized and turbulent ISM will be addressed in our 'EVLA cloud-scale survey of the local group galaxy M33' and in forthcoming surveys with the SKA.}

\FullConference{EXTRA-RADSUR2015 (*)\\
		20--23 October 2015\\
		Bologna, Italy

                \bigskip
                \hrule
                \bigskip

                \textnormal{(*) This conference has been organized
                  with the support of the Ministry of Foreign Affairs
                  and International Cooperation, Directorate General
                  for the Country Promotion (Bilateral Grant Agreement
                  ZA14GR02 - Mapping the Universe on the Pathway to
                  SKA)}
}

\begin{document}

\section{Introduction}
The most fundamental building blocks of galaxies, stars, are born within  clouds of gas and dust and during their lives they enrich the gas and the interstellar medium (ISM) with heavy elements, magnetic fields, and cosmic rays, all of which strongly affect the subsequent formation of stars and their host galaxy. Star formation feedback is important because it can suppress the formation of new stars by removing the surrounding gas via strong winds. On the other hand, it can trigger the formation of more stars, e.g., in gas condensed in supernova shells (e.g., Hensler 2010). The star formation feedback is important because it can re-distribute the gas and drive outflows in galaxies, changing the energetics of the ISM and the host galaxy.  To understand the evolution and appearance of galaxies it is therefore crucial to study the ISM/star formation interplay in galaxies. Most of our information about the ISM relates to its massive component, the gas and its various phases. Its interplay with  star formation  can be addressed through the famous Kennicutt-Schmidt relation between the rate of star formation and the gas density. However, not much is known about the most energetic ISM components, the cosmic rays and magnetic fields, and their role in structure formation in galaxies.

\subsection{Radio Continuum Emission: An Ideal SFR Tracer}

Radio continuum emission, as a tracer of the star formation rate (SFR), has been used  frequently in the literature ever since the discovery of its tight correlation with the infrared emission in galaxies (see Condon 2002, and references therein), even though some authors have raised the possibility of a conspiracy of several factors tightening the radio-IR correlation (Bell 2003, Lacki et al. 2003). 

Resolved observations of nearby galaxies show that the thermal and the nonthermal components of the radio continuum emission, mapped at $\gtrsim$ 200 pc linear resolutions, are both strong in star forming regions (Tabatabaei et al. 2007, 2013a, 2013b) . The separated nonthermal radio maps offer a direct basis for using the nonthermal radio continuum emission as a dust-unbiased SFR tracer in galaxies.  This is understandable as massive star formation activities such as supernova explosions, their shocks, and their remnants increase the number density of high-energy cosmic ray electrons (CREs) and/or accelerate them, on the one hand, and amplify the turbulent magnetic field strength, on the other. The net effect of these processes is strong nonthermal emission in or around star forming regions. Compared to the thermal emission, the contribution to the diffuse extended emission of the nonthermal component could be larger than that of the thermal component, depending on the diffusion of the cosmic ray electrons and the magnetic fields in the ISM (Tabatabaei et al. 2013b) . This leads to a sub-linear correlation between the nonthermal emission and the IR ( another SFR tracer) at sub-kpc resolutions, and a super-linear correlation globally (Heesen et al. 2014). Although our recent study shows no significant difference between the thermal and nonthermal emission when comparing them with hybrid SFR tracers such as H$\alpha$+24$\mu$m and FUV+ 24$\mu$m in the KINGFISH (Kennicutt et al. 2011) galaxy sample (Fig. 1, see Tabatabaei et al., in prep.).  Hence, globally, the nonthermal emission is a good SFR tracer, similar to the thermal emission. 

\begin{figure}
\begin{center}
\resizebox{\hsize}{!}{\includegraphics*{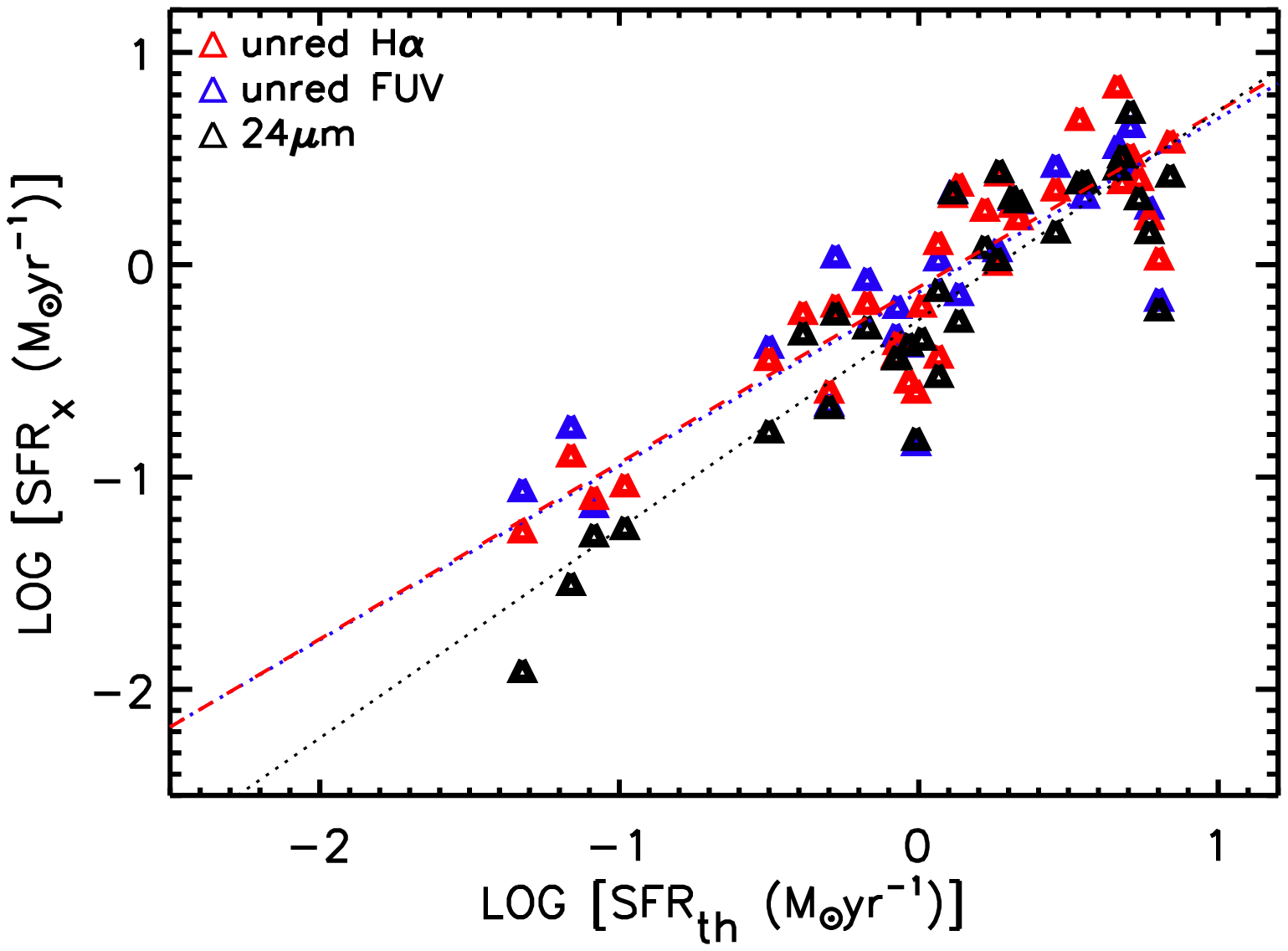}\includegraphics*{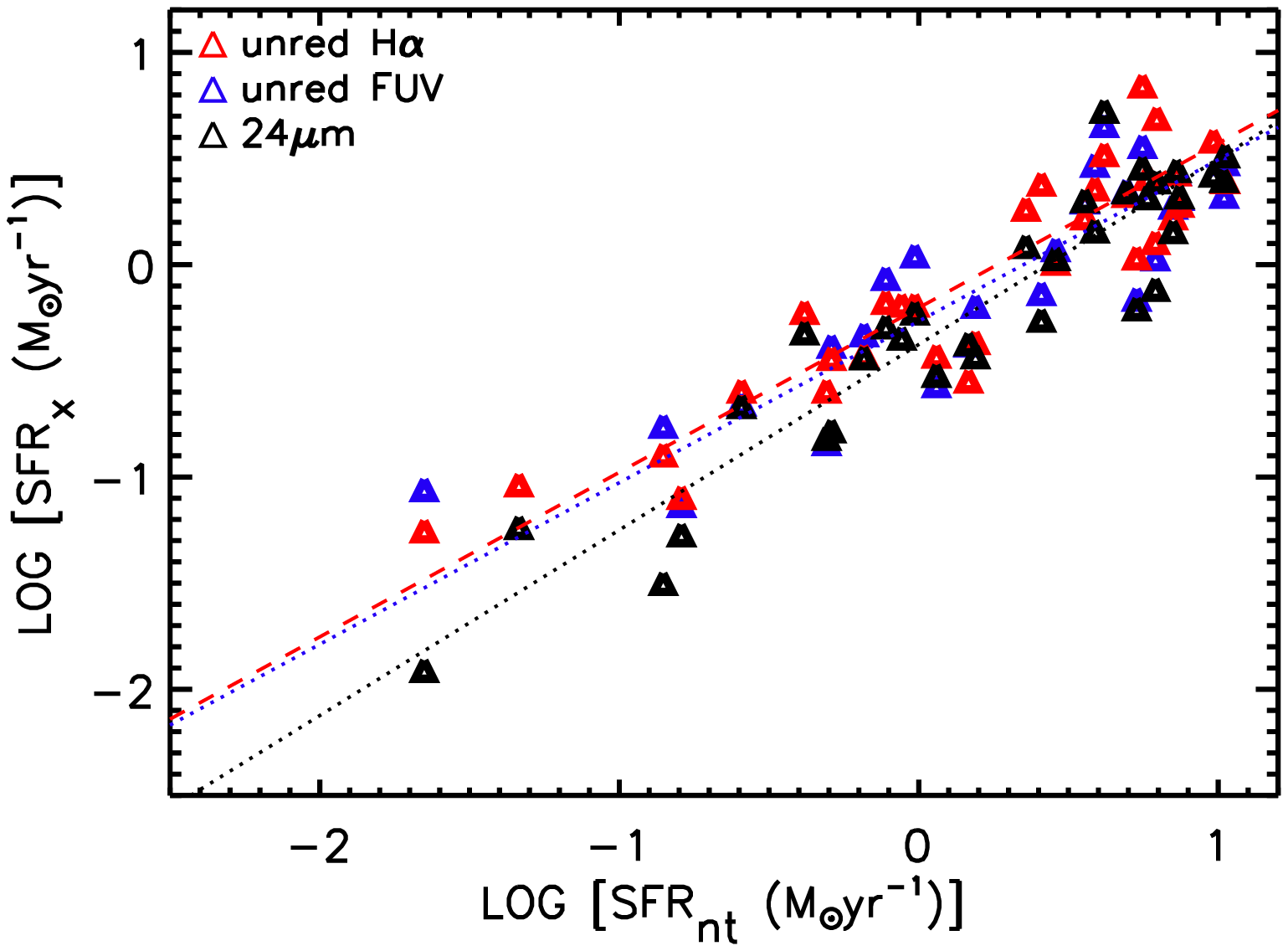}}
\caption[]{Star formation rates of the KINGFISH sample obtained using the H$\alpha$+24$\mu$m (red), FUV+24$\mu$m (blue), and 24$\mu$m (black) tracers against the thermal (left) and nonthermal (right) radio continuum star formation rate tracers. The radio SFR calibrations were obtained following Murphy et al. (2011).}
\label{fig:fig3}
\end{center}
\end{figure}

\section{Star Formation and Cosmic Rays}
Supernova explosions/remnants are considered to be the main source of cosmic rays. CREs can also be generated by the interaction of cosmic ray nucleons and ISM gas. These particles are energetically important, containing at least as much energy density as the interstellar gas and the magnetic fields. They must therefore play an important role in the regulation of star formation  (e.g., Jubelgas et al. 2008). CREs are expected to lose their energy at different rates depending on various cooling mechanisms in galaxies.  Observationally, this is confirmed by mapping the spectral index of the nonthermal radio emission (with a power-law spectrum) in M33 and NGC~6946 (Tabatabaei et al. 2007, 2013a). In star forming regions, the observed nonthermal spectrum (or the CRE energy index) is flat with a typical index of that expected theoretically for supernova remnants (~0.5-0.6). This means that CREs are more energetic in star forming regions due to recent injection or acceleration by strong shocks. Star formation could also influence the global spectrum of the CRE population in galaxies. The radio Spectral Energy Distribution (SED) survey of the KINGFISH galaxies indicates that the global nonthermal spectral index is flatter in more star forming galaxies (Tabatabaei et al., in prep.). 

\section{Star Formation and Magnetic Fields}
Star formation and feedback in galaxies cannot be understood when one neglects the connection between the magnetic fields and star formation. Radio synchrotron maps of  galaxies, along with their polarization data, have enabled us to  map the strength of the ordered, turbulent, and the total magnetic fields (Tabatabaei et al. 2008 and 2013a). We showed that the strength of the total and turbulent magnetic fields is correlated with the SFR surface density (Fig. 2). This correlation could be due to amplification of the turbulent magnetic field in star forming regions. Whether this enhanced magnetic field affects  the formation of new stars will be addressed in our M33 cloud-scale radio continuum survey, being performed with the Karl Jansky VLA in the C and L bands.
\begin{figure}
\begin{center}
\resizebox{9cm}{!}{\includegraphics*{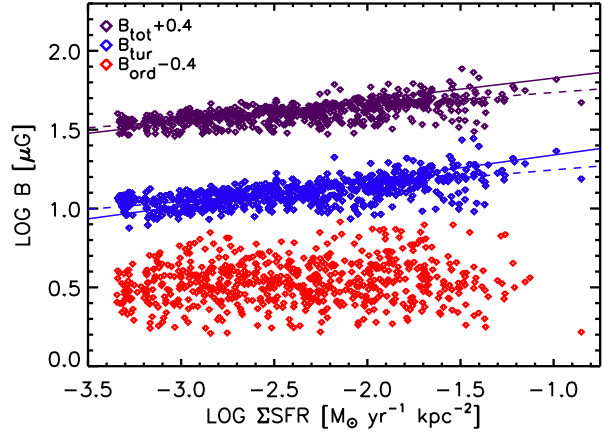}}
\caption[]{Resolved magnetic field strength vs. the star formation rate in NGC 6946 (Tabatabaei et al. 2013a). Blue  symbols show the turbulent magnetic field, red symbols the ordered magnetic field (shifted by -0.4 along the Y axis), and purple symbols the total magnetic field (shifted by +0.4 along the Y axis).  The ordered magnetic field does not seem to be correlated with the SFR, instead it is connected to the large-scale dynamics of galaxies (Tabatabaei et al. 2016). }
\label{fig:fig3}
\end{center}
\end{figure}
\section{Cloud-Scale EVLA Survey of Star Formation and Feedback in M33}
M33 is a Local Group galaxy with slightly sub-Solar metallicity, which makes it an ideal stepping stone to less regular and lower-metallicity objects such as dwarf galaxies and, probably, young-universe objects. We have performed a wide-band full-polarization mosaic of M33 with the EVLA in the 1-2 GHz L-band  covering a one square degree area around the galaxy center, and in the 4.5-6.5 GHz C-band covering the star forming disk (inner 20' x 20' disk) at ~10'' resolution, comparable to the resolution of the IRAM 30-m CO(2-1) data (Braine et al. 2010, Gratier et al. 2010). This resolution allows mapping the cosmic ray electron energy index at a level about 9 times better than before (Tabatabaei et al. 2007). Using this dataset and taking  full advantage of the Herschel M33 Extended Survey (HerM33es, Kramer et al. 2010) and Spitzer data, we uncover different phases of star formation from young stellar objects to supernovae, study the role of magnetic fields in molecular cloud formation and star formation, and dissect the star formation feedback on scales of giant molecular clouds (GMC) and larger. 

\subsection{Radio Continuum Emission \& GMC-Scale Star Formation Rate}
As the first step, we have investigated the use of the radio continuum emission as  an SFR tracer on GMC scales by comparing the EVLA radio data with the SFR traced using the MIPS 24$\mu$m data. An almost linear correlation holds between the SFR surface density and the 6 GHz radio luminosity density. The GMC-scale correlation agrees well with the global correlation found for the KINGFISH galaxy sample.

\begin{figure}
\begin{center}
\resizebox{9cm}{!}{\includegraphics*{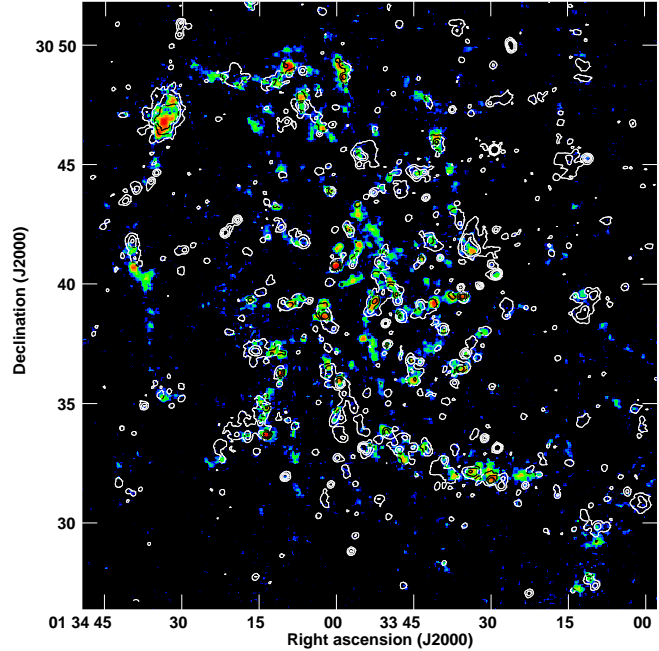}}
\caption[]{M33- The radio continuum emission at 6 GHz (contours) observed with the EVLA and the CO(2-1) line emission (color) at GMC-scale spatial resolution ($\sim$40 pc). }
\label{fig:fig3}
\end{center}
\end{figure}

\subsection{Radio-CO Correlation}
Comparing the maps of the radio continuum and the CO(2-1) line emission, we found  very good agreement (Fig. 3). The pixel-by-pixel correlation is characterized by a  Spearman rank coefficient of 0.75 $\pm$ 0.03 (Fig. 4). A tight radio-CO correlation was also found in other galaxies such as M51 (Schinnerer et al. 2013). This correlation could be due to the thermal free-free emission, if the neutral molecular gas is well mixed with the ionized gas or if their energy sources are the same.  The correlation could, on the other hand, be due to the nonthermal component of the radio continuum emission if a balance holds between the  cosmic ray/magnetic field and molecular gas pressures on scales of 40\,pc and larger. This is an important suggestion which, if true, would explain the unexpectedly tight radio-FIR correlation on small scales in this galaxy (Tabatabaei et al. 2013b, the radio-FIR correlation usually breaks down on a sub-kpc scale in galaxies). Moreover, it could also indicate the importance of the magnetic fields in the ISM structure formation with possible consequences for the formation of stars.  Separation of the thermal/nonthermal radio components is required to address this question. Wide-band observations at different radio bands are required to decompose the thermal/nonthermal components taking into account the possible curvatures in the nonthermal  SED.  The radio SED synthesis should be possible with our EVLA multi/wide-band observations and spectral synthesis. The thermal/nonthermal maps of M33 will be obtained just using the radio data and at about 9 times higher resolution than that presented in Tabatabaei et al. (2007). This will also allow identification of different phases of massive star formation (embedded young star clusters, HII regions, and SNRs) and their impact on the magnetized ISM. 

Such a resolved spectral synthesis will be routine with the SKA, even for distant galaxies, taking into account its frequency coverage from 70 MHz to 25 GHz in three bands (low: 70-450 MHz, mid: 0.3-10 GHz, and high: 5-25 GHz), sensitivity, survey speed, image fidelity, temporal resolution, and field of view (Garrett et al. 2010).
\begin{figure}
\begin{center}
\resizebox{8cm}{!}{\includegraphics*{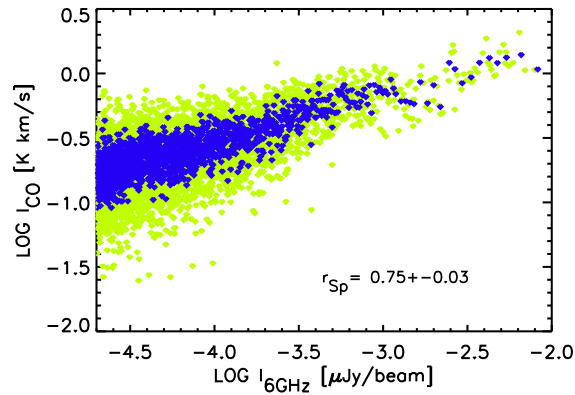}}
\caption[]{The pixel-by-pixel correlation between the molecular gas traced with the CO(2-1) emission (Fig.3) and the radio continuum emission at 6\,GHz. Blue points show the binned version of the data (green points).  }
\label{fig:fig3}
\end{center}
\end{figure}

\end{document}